\documentstyle[12pt]{article}

\begin{document}
\setcounter{page}{0}
\begin{titlepage}
\title{The Ising spin glass in a transverse field. Results
 of two fermionic models.}
\author{ Alba Theumann\\
	 Instituto de F\'{\i}sica\\ 
	 Universidade Federal do 
	 Rio Grande do Sul,\\
	 Av.  Bento Gon\c{c}alves 9500,
	 C. P. 15051\\
	 91501--970, Porto Alegre, RS, Brazil.\\[4mm] 
         A. A. Schmidt and S. G. Magalh\~aes\\
	 Departamento de Matem\'atica,\\
	 Universidade Federal de Santa Maria,\\
	 97119-900, Santa Maria, RS, Brazil.}
\date{}
\maketitle
\thispagestyle{empty}
\end{titlepage}

\begin{abstract}
\normalsize
\noindent
We analyze the long range Ising spin glass in a transverse field $\Gamma$ by using 
Grassmann variables in a field theory where the spin operators
are represented by bilinear combinations of fermionic fields. We compare the
results of two fermionic models. 
In the four state (4S)-model the diagonal $S_{i}^{z}$ operator has two vanishing 
eigenvalues, that are suppressed by a restraint in the two states (2S)-model. 
Within a replica symmetric theory and in the static approximation we obtain similar results 
for both models.
They both exhibit a critical temperature $T_{c}(\Gamma)$ that decreases when $\Gamma$
increases, until it reaches a quantum critical point (QCP) at the same value
of $\Gamma_{c}$ and
 they are both unstable under replica symmetry breaking in the whole spin glass phase.
\end{abstract}

\newpage
\section{Introduction}

The Ising model in a transverse field is widely studied for being the simplest system 
of interacting spins with quantum dynamics. The most striking feature is that 
the competition between thermal and quantum fluctuations reduce the critical temperature 
up to a point when a quantum phase transition occurs at $ T = 0 $, at a quantum critical 
point(QCP)\cite{1}. We will not discuss here the extensive literature on
results for several versions 
of the model, but we will concentrate instead in the quantum Ising spin glass in a transverse 
field. This is represented by a Hamiltonian in which only one component of the spins, 
say the z-component, interact among themselves with a random interaction while a uniform, 
constant field $ \Gamma$ is applied in the transverse x-direction.
The experimental realizations of this model are the $ LiHo_{x}Y_{1-x}F_{4}$ compounds\cite{2}.

In the calculation of the quantum mechanical partition function special tools are needed to 
deal with the non-commuting operators forming the Hamiltonian. The method more currently 
used in the study of short-range \cite{3} and infinite range \cite{4} spin glasses in a 
transverse field  is the Trotter-Suzuki formula \cite{5}, that maps a system of quantum spins 
in d-dimensions to a classical system of spins in (d + 1)-dimensions, and it is suited to 
perform numerical studies. Another way of dealing with the non-commutativity of quantum 
mechanical spin operators is to use Feynman's path integral formulations \cite{6,7} 
and to introduce 
time-ordering by means of an imaginary time $0\leq \tau\leq\beta$, where $\beta$ is the
inverse temperature. The work by Bray and Moore\cite{6} established
the basis for recent developments in the theory of the quantum Heisenberg spin glass\cite{8}.

A still different functional integral formulation consists in using
Grassmann variables to write a field theory with an effective action where the spin 
operators in the Hamiltonian are expressed as bilinear combinations of 
fermions\cite{9,10}.
The advantage of the fermionic formulation is that it has a natural application to
problems in condensed matter theory, where the fermion operators represent 
electrons that also participate in other physical processes, like 
superconductivity\cite{11,12} and the Kondo effect\cite{13}.
  In the present paper we use two fermionic models within a 
grassmannian field theory\cite{9}
to analyze the long range Ising spin glass in a transverse field.The novelty resides in the
method, as all previous results rely on the Trotter-Suzuki approximation\cite{4}.  
A  criticism to the fermionic formulation may be that the spin eigenstates at each site do not 
belong to one irreducible representation $ S^{z}=\pm\frac{1}{2}$, but they are labeled 
instead by the fermionic occupation numbers $n_{\sigma}= 0$ or 1 , giving two more  
states with $S^{z}=0$. We call this the "four states" (4S) model, and despite the 
presence of these two unwanted states the 4S-Ising spin glass model describes a spin glass 
transition with the same characteristics as 
the Sherrington - Kirkpatrick (SK) model \cite{14} in a replica symmetric theory.
A way to get rid of the unwanted states was introduced before 
by Wiethege and Sherrington\cite{15} for non-random 
interactions and it consists in fixing the occupation number $n_{i\uparrow }+ n_{i\downarrow}$ 
by means of an integral constraint at every site. We refer to this as the "two states" 
(2S)-Ising model. 

In sect. 2 we analyse the 4S-Ising and 2S-Ising spin glass models in a transverse field, 
within the static approximation in a replica symmetric theory . 
The static ansatz neglects time fluctuations and may be considered an approximation similar 
to mean field theory. Numerical Monte Carlo solutions of Bray and Moore's equations indicate
that the static approximation reproduces the correct results at finite temperatures\cite{16}.
 When $\Gamma=0$ the static approximation reproduces the exact results 
obtained by other methods, in particular for the 2S-Ising spin glass model we recover 
SK equations \cite{14}. The results in both models are very similar; they both exhibit a 
critical spin glass temperature $T_{c}(\Gamma)$ that decreases when the strength $\Gamma$ 
of the transverse field increases, until it reaches a quantum critical point(QCP) at 
$\Gamma _ {c}$, $T_{c}(\Gamma _ {c})=0$. The value of $\Gamma_{c}$ is the same for both 
models and the 4S-Ising and 2S-Ising models are identical close to the QCP.
We obtained for both models that the replica symmetric solution is unstable \cite{17} 
in the whole spin glass phase, in agreement with previous results with the 
Trotter-Suzuki method\cite{4}. We left sect. 3 for discussions.

\newpage

\section{The model and results}    

The Ising spin glass in a transverse field is represented by the Hamiltonian

\begin{eqnarray}
 H= -\sum_{ij}J_{ij}S^{z}_{i}S^{z}_{j}-2\Gamma\sum_{i}S^{x}_{i}
\label{2.1}
\end{eqnarray}
where the sum is over the N sites of a lattice and $J_{ij}$ is a random coupling 
among all pairs of spins, with gaussian probability distribution:

\begin{equation}
P(J_{ij})=e^{-J_{ij}^{2}N/16J^{2}}{\sqrt{N/16\pi J^{2}}}\;.
\label{2.2}
\end{equation}
 
The spin operators are represented by auxiliary fermions fields: 
\begin{eqnarray}
S^{z}_{i}=\frac{1}{2}[n_{i \uparrow}-n_{i \downarrow}]\nonumber\\
S^{x}_{i}=\frac{1}{2}[a_{i \uparrow}^{\dagger}a_{i \downarrow} + 
a_{i \downarrow}^{\dagger}a_{i \uparrow}]
\label{2.3}
\end{eqnarray}
where the $a_{i\sigma}^{\dagger}(a_{i \sigma})$ are creation (destruction) operators with 
fermion anticommutatiom rules and $\sigma = \uparrow$ or$ \downarrow$ indicates the 
spin projections. The number operators $n_{i \sigma}=a_{i\sigma}^{\dagger}a_{i\sigma}= 0$ 
or 1, then 
 $S_{i}^{z}$ in Eq.(\ref{2.3}) has two eigenvalues $\pm\frac{1}{2}$ 
corresponding to $n_{i\downarrow}= 1 - n_{i\uparrow}$, and two vanishing eigenvalues 
when $n_{i\downarrow}=n_{i\uparrow}$.

We shall use the Lagrangian path integral formulation 
in terms of anticommuting Grassmann fields described  in previous 
publications \cite{9,10}, so we avoid  giving repetitious details. 
We consider two models: the unrestrained, four states model that has been used previously 
\cite{9,11,12,13}, and also the two states model of Wiethege and Sherrington 
where the number operators satisfy the 
restraint $n_{i\uparrow}+n_{i\downarrow}=1$, what gives $S_{i}^{z}=\pm\frac{1}{2}$, 
at every site \cite{15}

The partition function in the 4S-model is given by

\begin{eqnarray}
Z_{4S} = T_{r }e^{-\beta H }
\label{2.4}
\end{eqnarray}  
while in the restrained model it takes the form:
\begin {eqnarray}
Z_{2S}=T_{r}[e^{-\beta H}\displaystyle\prod_{j}\delta(n_{j\uparrow}+n_{j\downarrow}-1)]
\label{2.5}
\end{eqnarray}
where $\beta=\frac{1}{T}$ is the inverse temperature.

By using the integral representation for the  Kronecker $\delta$-function:
\begin{eqnarray}
\delta(n_{j\uparrow}+n_{j\downarrow}-1)=\frac{1}{2\pi}
\displaystyle \int_{0}^{2\pi}dx_{j}e^{ix_{j}[n_{j\uparrow}+n_{j\downarrow}-1]}
\label{2.6}
\end{eqnarray}
we can express $Z_{4S}$ and $Z_{2S}$ in the compact functional integral form
\begin{eqnarray}
Z\{\mu\}= \int D(\varphi{\ast}\varphi)\prod_{j}\frac{1}{2\pi} 
\displaystyle \int_{0}^{2\pi}dx_{j}e^{-\mu_{j}}e^{A\{\mu\}}
\label{2.7}
\end {eqnarray}
where:
\begin{eqnarray}
A\{\mu\}&=&\int_{0}^{\beta} \{\displaystyle\sum_{j,\sigma}[\varphi_{j\sigma}^{\ast}(\tau)
\frac{d}{d\tau}\varphi_{j\sigma}(\tau)+ \mu_{j} \varphi_{j\sigma}^{\ast}(\tau) 
\varphi_{j\sigma}(\tau)]-\nonumber \\
        & & H(\varphi_{j \sigma}^{\ast}(\tau),\varphi_{j \sigma}(\tau)\}
\label{2.8}
\end{eqnarray}
and $\mu_{j}=0$ for the 4S-model while $\mu_{j}=ix_{j}$ for the 2S-model. 
Going to Fourier representation we introduce the spinors:
\begin{eqnarray}
\underline{\psi_{i}}(\omega)=\left(\begin{array}{c}\varphi_{i\uparrow}(\omega)\\
 \varphi_{i \downarrow}(\omega)\end{array}\right)
\label{2.9}
\end{eqnarray}
and the Pauli matrices:
\begin{eqnarray}
\underline{\sigma_{x}}=\left(\matrix{0&1\cr                         
				     1&0}\right)
\hskip 1cm
\underline{\sigma_{y}}=\left(\matrix{0&-i\cr                         
				     i&0}\right)
\hskip 1cm
\underline{\sigma_{z}}=\left(\matrix{1&0\cr                        
				     0&-1}\right)\;
\label{2.10}
\end{eqnarray}
to write the spin glass part of the action
\begin{equation}
A_{SG}=\displaystyle \sum_{\Omega}\sum_{i,j}\beta J_{ij}S_{i}^{z}(\Omega) S_{j}^{z}(\Omega)
\label{2.11}
\end{equation}
where
\begin{equation}
S_{i}^{z}(\Omega)= \frac{1}{2} \displaystyle \sum_{\omega}\underline{\psi}^{\dagger}(\omega + 
\Omega) \underline{\sigma}^{z}\underline{\psi}(\omega)
\label{2.12}
\end{equation}
with Matsubara's frequencias $\omega=(2n+1)\pi $ and $\Omega=2m\pi$ .
 In the static approximation, we retain just the term $\Omega=0$ in the sum over 
the frequency $\Omega$.

The transverse part of the action is given by:
\begin{equation}
A_{\Gamma}=\displaystyle\sum_{j} \underline {\psi}_{j}^{\dagger}(\omega)
\underline{\gamma}_{j}^{-1}(\omega)\psi_{j}(\omega)
\label{2.13}
\end{equation}
where the inverse propagator is
\begin {equation}
\underline{\gamma}_{j}^{-1}= i\omega +\mu_{j} +\Gamma \underline{\sigma}_{x}
\label{2.14}
\end{equation}
and the total action can be rebuild as 
\begin{equation}
A\{\mu\}=A_{\Gamma}+A_{SG}^{st}
\label{2.15}
\end{equation}
where $A_{SG}^{st}$ is the static component of Eq.(\ref{2.11}). 
We are now able to follow the standard procedures to get the configurational averaged 
free energy per site by using the replica formalism:
\begin{equation}
F=-\frac{1}{\beta N}\lim_{n\rightarrow 0}\frac{Z(n)-1}{n}
\label{2.16}
\end{equation}
where the configurational averaged, replicated, partition function $< Z^{n}>_{c,a}=Z(n)$ 
becomes, after averanging over $J_{ij}$:
\newpage
\begin{eqnarray}
Z(n)&=& \int_{-\infty}^{\infty}\prod_{\alpha \beta} dq_{\alpha \beta}
e^{-N \frac{\beta^{2}J^{2}}
{2}\displaystyle \sum_{\alpha \beta}q_{\alpha \beta}^{2}} \nonumber \\
    & &\prod_{j}\{\displaystyle \prod_{\alpha}\frac{1}{2\pi}\displaystyle
\int_{0}^{2\pi}dx_{j \alpha} e^{-\mu_{j \alpha}}\Lambda_{j}({q_{\alpha \beta}}) \}
\label{2.17}
\end{eqnarray}
with the replica index $\alpha=1,2,..,n$, and 
\begin{eqnarray}
\Lambda_{j}({q_{\alpha \beta}})&=& \int D(\varphi _{\alpha}^{\dagger} 
\varphi_{\alpha})exp[\sum_{\alpha}\displaystyle \sum_{\omega}\underline{\psi}^{\dagger \alpha}
(\omega) \underline{\gamma}_{j}^{-1}(\omega)\underline{\psi}^{\alpha}(\omega) 
+ \nonumber \\ & & \beta^{2}J^{2}4\displaystyle \sum_{\alpha \beta}q_{\alpha \beta}
S_{j}^{z \alpha}S_{j}^{z \beta}]
\label{2.18}
\end{eqnarray}

We indicate by $S_{j}^{z}$ the static component $ S_{j}^{z}(\Omega = 0)$ 
of Eq.(\ref{2.12}). 
We assume a replica symmetric solution of the saddle point equations:
\begin{eqnarray}
q_{\alpha \neq \beta}=q \hskip 2 cm q_{\alpha
\alpha}= q+\bar{\chi}
\label{2.19}
\end{eqnarray}
where q is the spin glass order parameter and $\bar{\chi}$ is related 
to the static susceptibility by $\bar{\chi}=T\chi$.

The sums over $\alpha $ in the spin part of the action produce again quadratic terms that can be linearized by introducting new auxiliary fields, with the result
\begin{eqnarray}
\Lambda_{j}(q,\bar{\chi}) = \int_{-\infty}^{\infty}Dz \prod _{\alpha}\displaystyle \int_{-\infty}^{\infty}D\xi_{\alpha} I_{j \alpha}(q,\bar{\chi},\mu_{j \alpha},z,\xi_{\alpha})
\label{2.20}
\end{eqnarray}
where $ Dy=\frac{1}{\sqrt{2\pi}}dy e^{-\frac{1}{2}y^{2}}$ and
\begin{equation}
I_{j \alpha}=\int D ( \varphi_{\alpha}^{\dagger}\varphi_{\alpha}) e^{\displaystyle\sum_{\omega}\underline{\psi}^{\dagger}(\omega)\underline{G}_{j}^{-1}(\omega)\underline{\psi}^{\alpha}(\omega)}
\label{2.21}
\end{equation}
with
\begin{equation}
\underline G_{j}^{-1}(\omega)=\underline\gamma_{j}^{-1}(\omega)+h_{\alpha}\underline\sigma_{z}
\label{2.22}
\end{equation}
\begin{equation}
h_{\alpha}=\beta J\sqrt{2q}z+\beta J\sqrt{2\bar{\chi}}\xi_{\alpha}
\label{2.23}
\end{equation}

The gaussian integral over Grassmann variables is straigthforward\cite{9,11,13}, 
giving the result:
\begin{equation}
\ln(I_{j\alpha})=\sum_{\omega}\ln[(i\omega+\mu_{j})^{2}-\Delta_{\alpha}]
\label{2.24}
\end{equation}
\begin{equation}
\Delta_{\alpha}=[\beta J \sqrt{2q}z + \beta J \sqrt{2\bar \chi}\xi_{\alpha}]^{2} 
+(\beta \Gamma)^{2}
\label{2.25}
\end{equation}

The sum over frequencias can be also easily performed\cite{9,11,13}  and we obtain
\begin{eqnarray}
I_{j\alpha}=1+e^{2\mu_{j \alpha}}+e^{\mu_{j\alpha}} 2 \cosh \sqrt{ \Delta_{ \alpha}}
\label{2.26}
\end{eqnarray}

From Eq.(\ref{2.17}), Eq.(\ref{2.20}) and Eq.(\ref{2.26}) we
 obtain at the saddle point:
\begin{eqnarray}
Z(n)&=&e^{-n N \frac{\beta^{2} J^{2}}{2}[\bar{\chi}^{2}+2 q \bar {\chi}]}\prod_{j} 
\{ \displaystyle \int_{- \infty}^{\infty}Dz \displaystyle \prod_{\alpha} 
\displaystyle \int_{-\infty}^{\infty}D \xi_{\alpha} \frac{1}{2 \pi} \nonumber \\
    & &\displaystyle \int_{0}^{2 \pi}dx_{j \alpha}[e^{- \mu_{j \alpha}}+
e^{\mu_{j \alpha}}+2 \cosh{\sqrt{\Delta_{\alpha}}}] \}
\label{2.27}
\end{eqnarray}

For the four states (4S) model there is no restraint and $\mu_{j \alpha}=0$, 
then the integrals over $x_{j \alpha}$ equal unity in Eq.(\ref{2.27}). 
For the restrained two states (2S) model we have $\mu_{i \alpha}=i x_{j \alpha}$ 
from Eq.(\ref{2.16}), then the integrals over the exponential terms identically 
vanish in Eq.(\ref{2.27}). We then obtain for the model with $ 2 (p+1)$ states, 
$p= 0$ or $1$:

\begin{equation}
\beta F_{p}=\frac{1}{2}(\beta J )^{2}[\bar{\chi}^{2}+ 2q \bar{\chi}]- 
\int_{-\infty}^{\infty}Dz \log [2 K_{p}(q,\bar{\chi},z)]
\label{2.28}
\end{equation}
where
\begin{equation}
K_{p}(q,\bar{\chi},z)=p +\int_{-\infty}^{\infty}D \xi \cosh \sqrt{\Delta}
\label{2.29}
\end{equation}

The saddle point equations for the order parameters are:
\begin{equation}
\overline{ \chi }_{p}= \int_{- \infty }^{ \infty } Dz
\frac{1}{K_{p}} \displaystyle \int_{-\infty }^{ \infty } D \xi  
\{ \frac{ h^{2} }{ \Delta } \cosh{ \sqrt{ \Delta }}+ 
\frac{ \beta^{2} \Gamma^{2} }{\Delta^{ \frac{3}{2} } }
 \sinh{ \sqrt{ \Delta }} \}- q_{p} 
\label{2.30}
\end{equation}
\begin{equation}
q_{p}=\int_{-\infty}^{\infty}Dz
\frac{1}{K_{p}^{2}} \{ \displaystyle \int_{-\infty}^{\infty}D \xi \frac {h}{\sqrt{\Delta}}
 \sinh{ \sqrt{ \Delta }} \}^{2} 
\label{2.31}
\end{equation}
where $h(\xi,z)$ is given in Eq.(\ref{2.23}). We obtain for the de Almeida-Thouless eigenvalue 
\cite{17} and entropy in both models:
\begin{eqnarray}
\lambda_{p}^{AT}&=&1-2(\beta J)^{2} \int_{-\infty }^{ \infty } Dz \frac{1}{ K_{p}^{4}} 
\{ K_{p} \displaystyle \int_{-\infty}^{\infty } D \xi [\frac{ h^{2} }{ \Delta } 
\cosh{ \sqrt{ \Delta }}+\nonumber \\
                & & \frac{(\beta \Gamma )^{2}}{ \Delta^{ \frac{3}{2} } } 
\sinh{ \sqrt{\Delta}}] -[\displaystyle \int_{-\infty}^{\infty}D \xi\frac{\lambda}
{\sqrt{\Delta}}\sinh \sqrt{\Delta}]^{2} \} 
\label{2.32}
\end{eqnarray}
\begin{eqnarray}
\frac{S}{K}&=&- \frac{3}{2}(\beta J)^{2} ( \bar{\chi}^{2}+2 \bar{\chi}q) +
 \int_{-\infty}^{\infty} Dz \log{(2 K_{p})}-\nonumber\\
           & &(\beta \Gamma)^{2} \displaystyle \int_{-\infty}^{\infty}Dz
\frac{1}{K_{p}}\displaystyle \int_{-\infty}^{\infty}D\xi 
\frac{\sinh{\sqrt\Delta}}{\sqrt{\Delta}}
\label{2.33}
\end{eqnarray}

The Landau expansion of the free energy in powers of q gives:
\begin{equation}
\beta F_{p}= \beta F_{p}^{0}+B_{p}q^{2}+C_{p}q^{3}
\label{2.34}
\end{equation}
where the coefficients are:
\begin{eqnarray}
B_{p}=[D_{p}-1][D_{p}-2]\nonumber\\
C_{p}=- \frac{4}{3}[D_{p}-1]\{2(D_{p}-1)^{2}+ 3( D_{p}-2)\}
\label{2.35}
\end{eqnarray}
and
\begin{equation}
D_{p}=\frac{p+J_{1}}{p+J_{0}}
\label{2.36}
\end{equation}
\begin{equation}
J_{\l}=\int_{-\infty}^{\infty}D\xi  \xi^{2\l}
\cosh{\sqrt{2 \bar{\chi}_{0}\beta^{2}J^{2}\xi^{2}+\beta^{2}\Gamma^{2}}}
\label{2.37}
\end{equation}

In Eq.(\ref{2.37}) we need also:
\begin{equation}
\bar{\chi}(q=0)=\bar{\chi}_{0}=\frac{1}{\sqrt{2}\beta J}\sqrt{D_{p}-1}
\label{2.38}
\end{equation}

As $C_{p}<0$, the spin glass phase is characterized by $B_{p}>0$, giving a maximum 
instead of a minimum \cite{14} of the free energy. The critical temperature is 
obtained by solving simultaneously:
\begin{eqnarray}
D_{p}=\frac{p+J_{1}(\beta_{c})}{p+J_{0}(\beta_{c})}=2 \nonumber\\
\bar {\chi}_{0}=\frac{1}{\sqrt{2}\beta_{c}J}
\label{2.39}
\end{eqnarray}

The numerical results for the critical temperature $T_{c}(\Gamma)$ and the entropy 
$S_{0}=S(T_{c},\Gamma)$ are shown in Fig. 1 for the 4S-model and Fig. 2 for the 2S-model. 
For large values of the transverse field $\Gamma$ the 2S-model and 4S-model are 
undistinguishable. The analytic soluction of Eq.(\ref{2.39}) when $T_{c}=0$ 
gives the critical value $\Gamma{c}=2\sqrt{2}J$ for both models. When $\Gamma = 0$, 
the equations (30)-(33) for the 2S-model (p=0) reproduce the Sherrington-Kirkpatrick 
\cite{14} results, while for the 4S-model (p=1), we recover our previous results \cite{9}.

Finally, we comment on the de Almeida-Thouless instability. The exact soluction 
for $\lambda^{AT}$ in Eq.(\ref{2.32}) in both limits, $\Gamma=0$ and $\Gamma = \Gamma_{c}$, 
shows that $\lambda^{AT}=0$ at the transition point both for the 2S-model and the 4S-model, 
while numerical results confirm that $\lambda^{AT}(T_{c})=0$ for both models on the 
critical line $T_{c}(\Gamma)$. This is a correct result and A.T. acknowledges a flaw in 
a previous publication \cite{18}.
\newpage
\section{Discussion}

We performed a new study of two quantum Ising spin glass models in a transverse field 
by means of a 
path integral formalism where the spin operators are represented by bilinear combinations 
of fermionic fields. All previous results in this problem were obtained with
the Trotter-Suzuki approximation\cite{4}. 
In the unrestricted four-states (4S)-model the fermionic representation gives for the   
diagonal $S_{i}^{z}$-operator  
two eigenvalues $S_{i}^{z}= \pm \frac{1}{2}$ and two vanishing eigenvalues, 
while in the state (2S)-model the vanishing eigenvalues are suppressed by means of an 
integral constraint. 
The results in both models were obtained with the static approximation and the phase 
diagram coincides with previous results with the Trotter-Suzuki method \cite{4}. 
Regarding the de Almeida-Thouless instability\cite{17}, we obtained that the replica symmetric 
solution is unstable in the whole spin glass phase.

In future work we will apply
the fermionic representation of the 
transverse Ising spin glass to problems in 
condensed mather theory and also the replica symmetry breaking in the ordered state
will be investigated.
\section{Acknowledgement} 
 
We acknowledge partial financial support from the Conselho 
Nacional de Desenvolvimento
Cient\'{\i}fico e Tecnol\'ogico (CNPq),Financiadora de Estudos 
e Projetos (Finep) and 
Funda\c{c}\~ao de Amparo \'a Pesquisa do Estado de Rio Grande 
do Sul (FAPERGS).
\newpage
\section{Figure Captions}
Fig1. Critical temperature $T_{c}(\Gamma)$ and 
entropy on the critical line $S_{c}(T_{c},\Gamma)$ for the
4S-model.\\

Fig2.  Same as Fig1. for the 2S-model

\newpage
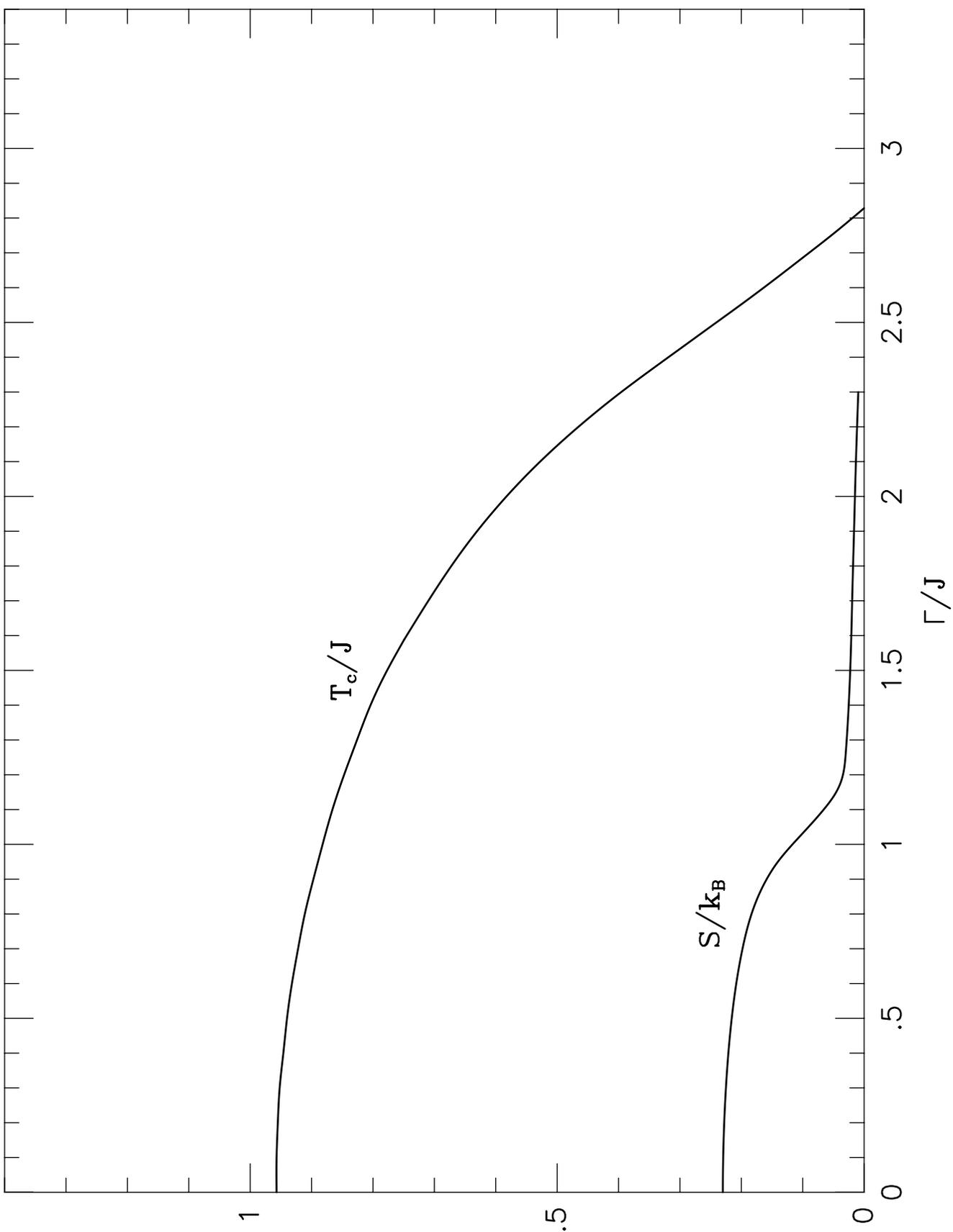
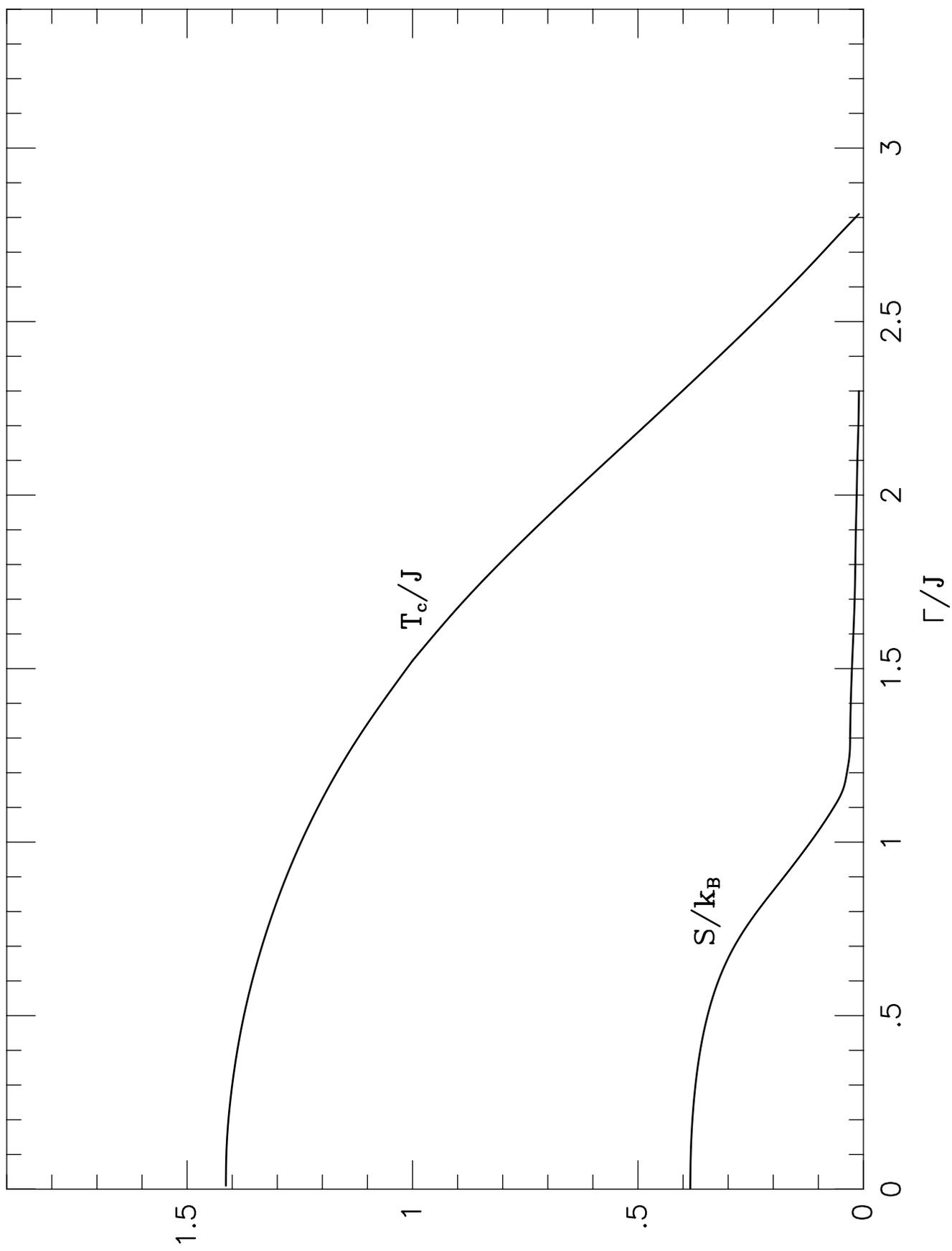
\end{document}